\documentclass[10pt]{article}
\usepackage[tbtags]{amsmath}
\usepackage{accents}
\usepackage{rotating}
\usepackage{amsfonts}
\usepackage{amssymb}
\usepackage{latexsym}

\DeclareMathSymbol{\vecarrow}{\mathord}{letters}{"7E}

\newcommand{\iu}{{\mathrm{i}}}

\newcommand{\fett}[1]{\mbox{\boldmath$#1$}}

\newcommand{\starMP}{*_{\scriptscriptstyle MC}}
\newcommand{\starM}{*_{\scriptscriptstyle M}}
\newcommand{\starP}{*_{\scriptscriptstyle C}}

\newcommand{\beq}{\begin{equation}}
\newcommand{\eeq}{\end{equation}}

\begin{document}
\makeatletter

\title{Star Products for Relativistic Quantum Mechanics}
\author{Peter Henselder\footnote{peter.henselder@uni-dortmund.de} \\
Fachbereich Physik, Universit\"at Dortmund\\
44221 Dortmund, Germany}
\maketitle

\begin{abstract}
The star product formalism has proved to be an alternative
formulation for nonrelativistic quantum mechanics. We want to
introduce here a covariant star product in order to extend the
star product formalism to relativistic quantum mechanics in the
proper time formulation.

\end{abstract}

\section{Introduction}
\qquad In deformation quantization the noncommutativity of 
quantum theory is described by introducing 
a noncommutative product for functions on the phase space. 
This star product emulates the product of operators in the 
conventional approach to quantum
mechanics. The star product is parametrized by a deformation
parameter $\hbar$ and in the limit $\hbar\rightarrow 0$ it goes
over to the pointwise product of phase space functions. In this
description quantum mechanics is then a deformed version of
classical mechanics, where the mathematical concept of deformation
goes back on Gerstenhaber \cite{Gerstenhaber}. The
advantage of doing quantum mechanics this way is that the
classical limit and the correspondence principle have a clear
meaning. Moreover in deformation quantization there is no severe
conceptual and formal break if one goes over from classical to
quantum mechanics.

Deformation quantization was first formulated as an autonomous 
approach to nonrelativistic quantum mechanics by Bayen et al.\ 
in \cite{Bayen}. The next step was then to
include spin in this formalism. In \cite{Deform3} and \cite{Deform5} 
this was done using the description of spin in the context of
grassmannian mechanics \cite{Berezin1}. Deformation quantization
of the simplest grassmannian system leads there to a grassmannian
star product and a description of spin. Moreover it appeared that
this grassmannian star product is nothing else then the Clifford
product that one knows from the superanalytic formulation of
geometric algebra \cite{DoranGull} (for 
a comprehensive review of geometric algebra see \cite{Doran1}). 
This allows to formulate geometric algebra as a
fermionic deformed version of superanalysis \cite{Deform6}. The relativistic
version of geometric algebra, i.e.\ space time algebra
\cite{Hestenes1}, was used to clarify the geometric setting of
Dirac theory (for a throughout discussion of Dirac theory 
in geometric algebra see \cite{Doran}). Formulation of space time
algebra with a fermionic star product leads then to a
description of Dirac theory in the context of deformation
quantization \cite{Deform5,Deform6}.

Nevertheless there are two problems that appear when one does Dirac
theory in the star product formalism. The first problem is that
the star product used to describe Dirac theory consists actually
of a three dimensional bosonic Moyal product and a four
dimensional fermionic Clifford product \cite{Deform5}. But with 
a three dimensional Moyal product it is not possible to describe
Lorentz transformations. Using a three dimensional Moyal star 
product is just a reflection of the special role that time plays in quantum
mechanics. The time variable is not as the space variables a phase
space coordinate that becomes in conventional quantum mechanics an
operator and in the star product formalism appears in the star
product. Time is rather used as the development parameter of the
Hamiltonian system, which can be seen as the reason for many
problems in Dirac theory \cite{Fanchi1}. The second problem that arises 
if one formulates Dirac theory in the context of deformation quantization
is the problem of the classical limit. In what sence Dirac theory
has a classical limit was for example discussed in \cite{Bolivar,Liang}.
But in deformation quantization, where the quantum theory is just a 
deformed version of a classical theory the classical theory has to be
directly reobtained in the limit $\hbar\rightarrow 0$. So a consistent
approach to relativistic quantum mechanics in deformation
quantization would be to start with a manifest covariant classical
hamiltonian system and then deform this system with a covariant
star product.

Both problems are solved if one uses the parametrized or proper
time formalism (for a throughout discussion see for example 
\cite{Fanchi2,Fanchi3,Johns}), where one distinguishes 
between time as a space time coordinate and the evolution parameter 
of the system. Moreover in the star product formalism one obtains
a formal unification of the geometric structure of space time algebra 
and the quantum structures in the proper time formalism. While in this 
paper the philosophy of deformation quantization leads to
a connection of geometric algebra and the proper time formalism there 
was also another approach by Pavsic to unify these two formalisms 
\cite{Pavsic}.

In the next section it will be shortly described how the star product 
formalism can be applied to Dirac theory. Therefore one uses the algebra morphism
of the operator and the star product algebra and the algebra morphism
of the algebra of the gamma matrices and the star product spacetime algebra. 
Section three shows then how one can use a four dimensional Moyal product to
describe Lorentz transformations just in the same way as in space time algebra,
so that the four dimensional Moyal and the four dimensional Clifford star products are
supersymmetric counterparts. The four dimensional Moyal product constructed
in order to obtain a star product formulation of the Loretz transformation is then a 
deformed product on an eight dimensional phase space, where time and energy
are the additional coordinates. This extended phase space is the arena for
parametrized relativistic classical mechanics, that is described in section four. 
One can then use the four dimensional Moyal product for deformation quantization
of this parametrized relativistic classical mechanics which is done in section five.
Furthermore the noncommutativity of the Moyal product leads in combination with
space time algebra to a natural appearance of a spin term. Such additional spin terms
that appear due to noncommutativity were also found in the nonrelativistic case
\cite{Deform6} and in the case of supersymmetric quantum mechanics \cite{Susy}.  

\section{Star Products in Dirac Theory}
\setcounter{equation}{0}
The algebraic structures of Dirac theory can be
described with a bosonic and a fermionic star product. On the one
hand one has the structures of space-time algebra in its
superanalytic formulation, i.e.\ the basis vectors of space-time
are the four Grassmann variables $\fett{\gamma}_{\mu}$, so that a
general space-time multivector has the form
\begin{equation}
A=A_{(0)}+A_{(1)}^{\mu}\fett{\gamma}_{\mu}
+A_{(2)}^{\mu\nu}\fett{\gamma}_{\mu}\fett{\gamma}_{\nu}
+A_{(3)}^{\mu\nu\rho}\fett{\gamma}_{\mu}\fett{\gamma}_{\nu}\fett{\gamma}_{\rho}
+A_{(4)}\fett{\gamma}_{0}\fett{\gamma}_{1}
\fett{\gamma}_{2}\fett{\gamma}_{3}.
\end{equation}
A four-vector is then a general Grassmann number of grade one:
$\fett{x}=x^{\mu}\fett{\gamma}_{\mu}$. Furthermore one has the
fermionic Clifford star product
\begin{equation}
A\starP B=A\, \exp\left[\eta_{\mu\nu}
\frac{\overleftarrow{\partial}}{\partial \fett{\gamma}_{\mu}}
\frac{\overrightarrow{\partial}}{\partial\fett{\gamma}_{\nu}} \right]\, B,
\label{Cliffpro}
\end{equation}
where $\eta_{\mu\nu}$ is the Lorentz metric. The Clifford star
product deforms the Grassmann algebra of the $\fett{\gamma}_{\mu}$
into a Clifford algebra. The scalar product of two vectors
$\fett{a}=a^{\mu}\fett{\gamma}_{\mu}$ and $\fett{b}=b^{\mu}
\fett{\gamma}_{\mu}$ is the scalar part of their Clifford product:
$\fett{a}\cdot\fett{b} =\langle\fett{a}\starP\fett{b}\rangle_0
=a^{\mu}b_{\mu}$, where $\langle\;\rangle_n$ projects on the term
of Grassmann grade $n$. This can also be written with the star
anticommutator $\{A,B\}_{\starP}=A\starP B+B\starP A$, so that one
has for two basis vectors
\begin{equation}
\{\fett{\gamma}_{\mu},\fett{\gamma}_{\nu}\}_{\starP}
=2\fett{\gamma}_{\mu}\cdot\fett{\gamma}_{\nu}=2\eta_{\mu\nu}.
\end{equation}

While the fermionic sector describes the geometric structure one
can then introduce noncommutativity by demanding that the scalar
multivector coefficients have to be multiplied by the bosonic
Moyal product
\begin{equation}
f\starM g=f\, \exp\left[\frac{\iu\hbar}{2}\sum_{i=1}^3
\left(\frac{\overleftarrow{\partial}}{\partial q^{i}}
\frac{\overrightarrow{\partial}}{\partial p_{i}}
-\frac{\overleftarrow{\partial}}{\partial p_{i}}
\frac{\overrightarrow{\partial}}{\partial q^{i}}\right)\right]\, g.
\end{equation}
It is then straight forward to describe Dirac theory in star
product formalism. Starting point is the Dirac Hamiltonian
\begin{equation}
H_D=\alpha_i p^i+\fett{\beta}m,
\end{equation}
with $\fett{\beta}=\fett{\gamma}_0$ and $\alpha_i
=\fett{\gamma}_0\starP\fett{\gamma}_i=\fett{\gamma}_0\fett{\gamma}_i$.
The vector $\fett{\beta}$ and the bivectors $\alpha_i$ fulfill for
the standard Lorentz metric
\begin{equation}
\fett{\beta}\starP\fett{\beta}=\alpha_i\starP\alpha_i=1, \qquad
\{\fett{\beta},\alpha_i\}_{\starP}=0\qquad\mathrm{and}\qquad
\{\alpha_i,\alpha_j\}_{\starP}=2\delta_{ij},
\end{equation}
so that $H_D\starMP H_D=\vec{p}^{\; 2}+m^2$. The next step is to
calculate the star exponential \cite{Bayen} as
\begin{equation}
e_{\starMP}^{\frac{1}{\iu\hbar}H_Dt}=
\sum_{n=0}^{\infty}\frac{1}{n!}\left(\frac{t}{\iu\hbar} \right)^n
H_D^{n\starMP} = \pi_{-E}^{(MP)}(\vec{p}\,)\,e^{+\iu tE/\hbar}
+\pi_{+E}^{(MP)}(\vec{p}\,)\,e^{-\iu tE/\hbar},
\end{equation}
with the Wigner functions
\begin{equation}
\pi_{\pm E}^{(MC)}(\vec{p}\,)= \frac{1}{2}\left(1 \pm\frac{H_D}{E}
\right) \label{diracpmEpi}
\end{equation}
and $E=\sqrt{\vec{p}^{\; 2}+m^2}$. The energy projectors $\pi_{\pm
E}^{(MP)}(\vec{p}\,)$ are idempotent, complete and fulfill the
$*$-eigenvalue equations
\begin{equation}
H_D\starMP\pi_{\pm E}^{(MP)}(\vec{p}\,)= \pm E\,\pi_{\pm
E}^{(MP)}(\vec{p}\,). \label{ham}
\end{equation}

For a unit vector $\vec{u}$ orthogonal to $\vec{p}$ one can find
projectors that are also $*$-eigenfunctions of the spin, which is
defined as $S_{\vec{u}}=\frac{\hbar}{2}\gamma^5
\starP(\fett{\gamma}_i u^i)$ with $\gamma^5=\iu\fett{\gamma}_0
\fett{\gamma}_1\fett{\gamma}_2\fett{\gamma}_3$. With $S_{\vec{u}}
\starP S_{\vec{u}}=\left(\frac{\hbar}{2}\right)^2$ the star
exponential is
\begin{equation}
e_{\starP}^{\frac{1}{\iu\hbar}S_{\vec{u}}\varphi}=
\sum_{n=0}^{\infty}\frac{1}{n!}\left(\frac{\varphi}{\iu\hbar}
\right)^n S_{\vec{u}}^{n\starP} =
\pi_{-\frac{1}{2}}^{(C)}(\vec{u}\,)\,e^{+\iu\varphi/2}
+\pi_{+\frac{1}{2}}^{(C)}(\vec{u}\,)\,e^{-\iu\varphi/2}
\end{equation}
with the Wigner functions
\begin{equation}
\pi_{\pm\frac{1}{2}}^{(C)}(\vec{u}\,)= \frac{1}{2}\pm
\frac{1}{\hbar} S_{\vec{u}}.\label{Susol}
\end{equation}
These are the star product analogues of the Dirac spin projectors
and they obey the $*$-eigenvalue equation
\begin{equation}
S_{\vec{u}}\starP\pi_{\pm\frac{1}{2}}^{(C)}(\vec{u}\,) =
\pm\frac{\hbar}{2}\pi_{\pm\frac{1}{2}}^{(C)}(\vec{u}\,).\label{Sugl}
\end{equation}

Since $H_D$ and $S_{\vec{u}}$ commute, i.e.\
$\left[H_D,S_{\vec{u}}\right]_{\starP}=H_D\starP
S_{\vec{u}}-S_{\vec{u}} \starP H_D=0$, the Wigner functions
$\pi_{\pm E}^{(MC)}(\vec{p}\,)$ and
$\pi_{\pm\frac{1}{2}}^{(C)}(\vec{u}\,)$ also commute and the
complete and idempotent Wigner functions for the Dirac problem are
given by
\begin{equation}
\pi_{\pm E,\pm\frac{1}{2}}^{(MC)}(\vec{p},\vec{u}\,)=\pi_{\pm
E}^{(MC)}(\vec{p}\,) \starMP
\pi_{\pm\frac{1}{2}}^{(C)}(\vec{u}\,). \label{pidi}
\end{equation}
The $*$-eigenvalue equations are then
\begin{equation}
H_D\starMP\pi_{\pm E,\pm\frac{1}{2}}^{(MC)}(\vec{p},\vec{u}\,)=
\pm E\pi_{\pm E,\pm\frac{1}{2}}^{(MC)}(\vec{p},\vec{u}\,)
\qquad\text{and}\qquad S_{\vec{u}}\starMP\pi_{\pm E,\pm
\frac{1}{2}}^{(MC)}(\vec{p},\vec{u}\,) = \pm\frac{\hbar}{2}
\pi_{\pm E,\pm\frac{1}{2}}^{(MC)}(\vec{p},\vec{u}\,).
\label{HSstargenvalue}
\end{equation}

\section{Active and Passive Lorentz Transformations}
\setcounter{equation}{0}
While it is formally possible to describe Dirac theory in the star
product formalism, there are two severe conceptual problems. First
of all Dirac theory has no classical limit. This contradicts the
philosophy of deformation quantization, which states that quantum
mechanics is obtained by deformation of classical mechanics and
setting the deformation parameter $\hbar$ to zero one reobtains
classical mechanics. So an approach to relativistic quantum
mechanics in the spirit of deformation quantization would be to
start with the manifest covariant version of Hamiltonian mechanics
and then deform this version of classical mechanics in a covariant
fashion. This approach will be described in the following section.

Closely related to this problem is the problem that the star
product formalism used in the last section is not supersymmetric,
i.e.\ one uses a four dimensional fermionic Clifford star product
but a three dimensional bosonic Moyal star product. This means
that it is not possible to describe Lorentz transformations in an
active and a passive manner. A passive Lorentz transformation is a
transformation of the basis vectors $\fett{\gamma}_{\mu}$. As a
rotation in space-time such a Lorentz transformation is generated
by the six space-time bivectors that can be written as
\begin{equation}
\sigma_{\mu\nu}=\frac{I_{(4)}}{2}\starP\left[\fett{\gamma}_{\mu},
\fett{\gamma}_{\nu}\right]_{\starP},
\end{equation}
where $I_{(4)}=\fett{\gamma}_0\fett{\gamma}_1\fett{\gamma}_2
\fett{\gamma}_3$ is the pseudoscalar. The generators for the
passive boosts and rotations are
\begin{equation}
\mathtt{K}_i=\frac{1}{2}\sigma_{0i}\qquad\mathrm{and}\qquad
\mathtt{L}_i=\frac{1}{2}\sum_{j<k}\varepsilon_{ijk}\sigma_{jk}
\end{equation}
and they satisfy in the case of the nonstandard metric (for the
standard metric one has to replace $I_{(4)}$ by $-I_{(4)}$):
\begin{equation}
\left[\mathtt{L}_i,\mathtt{L}_j\right]_{\starP}
=-I_{(4)}\starP\varepsilon_{ijk}\mathtt{L}_k,\qquad
\left[\mathtt{L}_i,\mathtt{K}_j\right]_{\starP}
=-I_{(4)}\starP\varepsilon_{ijk}\mathtt{K}_k,\quad\mathrm{and}\quad
\left[\mathtt{K}_i,\mathtt{K}_j\right]_{\starP}
=I_{(4)}\starP\varepsilon_{ijk}\mathtt{L}_k.
\label{passiveLorentzalg}
\end{equation}
The passive Lorentz transformation is then given by
\begin{equation}
\fett{q}'=e_{\starP}^{\frac{1}{4}I_{(4)}\starP\alpha^{\mu\nu}\sigma_{\mu\nu}}
\starP\fett{q}\starP
e_{\starP}^{-\frac{1}{4}I_{(4)}\starP\alpha^{\mu\nu}\sigma_{\mu\nu}}
=q^{\mu}\big(\Lambda_{\mu}^{\nu}\fett{\gamma}_{\nu}\big)
\end{equation}
where $\Lambda_{\nu}^{\mu}$ is the well known Lorentz
transformation matrix. In Dirac theory the passive transformations
are constructed a posteriori by demanding the invariance of the
four-vector $p_{\mu}\fett{\gamma}^{\mu}$, just as the basis
vectors of space-time are discovered a posteriori in a tuple
notation by factorizing the Klein-Gordon equation.

An active Lorentz transformation acts on the coefficients of the
four vector. Such an active Lorentz transformation can also be
described in the star product formalism. But one needs the four
dimensional Moyal product
\begin{equation}
f\,\starM\, g=f\,\exp\left[\frac{\iu\hbar}{2}\eta^{\mu\nu}
\left(\frac{\overleftarrow{\partial}}{\partial q^{\mu}}
\frac{\overrightarrow{\partial}}{\partial p^{\nu}}
-\frac{\overleftarrow{\partial}}{\partial p^{\mu}}
\frac{\overrightarrow{\partial}}{\partial q^{\nu}}\right)\right]\,g,
\label{comopro}
\end{equation}
where the nonstandard metric should be chosen, so that the three
dimensional part reduces to the conventional Moyal product. The
generators of an active Lorentz transformation are
\begin{equation}
M^{\mu\nu}=q^{\mu}p^{\nu}-p^{\mu}q^{\nu},
\end{equation}
with
\begin{equation}
\left[M^{\mu\nu},M^{\rho\sigma}\right]_{\starM}=\iu\hbar\left(
\eta^{\mu\rho}M^{\nu\sigma}-\eta^{\nu\rho}M^{\mu\sigma}
+\eta^{\mu\sigma}M^{\rho\nu}-\eta^{\nu\sigma}M^{\rho\mu}\right).
\end{equation}
The generators of boosts and rotations are
\begin{equation}
K^i=M^{01}\qquad\mathrm{and}\qquad
L^i=\sum_{j<k}\varepsilon^{ijk}M^{jk}.
\end{equation}
They form the following active Moyal star-commutator algebra
\begin{equation}
\left[L^i,L^j\right]_{\starM}=\iu\hbar\varepsilon^{ijk}L^k,\qquad
\left[L^i,K^j\right]_{\starM}=\iu\hbar\varepsilon^{ijk}K^k
\qquad\mathrm{and}\qquad
\left[K^i,K^j\right]_{\starM}=-\iu\hbar\varepsilon^{ijk}L^k,
\label{activeLorentzalg}
\end{equation}
so that an active Lorentz transformation of the four-vector
$\fett{q}=q^{\mu}\fett{\gamma}_{\mu}$ is given by
\begin{equation}
\fett{q}'=e_{\starM}^{-\frac{\iu}{\hbar}\alpha_{\mu\nu}M^{\mu\nu}}
\starM \fett{q}\starM
e_{\starM}^{\frac{\iu}{\hbar}\alpha_{\mu\nu}M^{\mu\nu}}
=\big(\Lambda^{\mu}_{\nu}q^{\nu}\big)\fett{\gamma}_{\mu}.
\end{equation}
Taking the translations with the generators $p_{\mu}$ into account
the Lorentz algebra is with
$\left[p_{\mu},p_{\nu}\right]_{\starM}=0$ and
\begin{equation}
\left[M^{\mu\nu},p_{\rho}\right]_{\starM}
=\iu\hbar\left(\eta^{\mu\rho}p_{\nu}-\eta^{\nu\rho}p_{\mu}\right)
\end{equation}
extended to the Poincar\'{e} algebra.

Using now the four dimensional Moyal product (\ref{comopro}) for
deformation quantization means that the one particle phase space
is extended by the two variables $q^0$ and $p^0$, which means that
the time development is not described by the time, that is now a
phase space coordinate but by an additional parameter. So what is
actually deformed by the four dimensional Moyal product
(\ref{comopro}) is parametrized Hamiltonian dynamics. And taking
the limit $\hbar\rightarrow 0$ the star product reduces to the
conventional product so that one obtains the classical undeformed
parametrized Hamiltonian dynamics, so that the conceptual problem
of the missing classical limit is also solved. In the operator
formalism of canonical quantization this would mean that time is
no longer a scalar but an operator, for a discussion 
concerning the existence of such a time operator see \cite{Galapon}.

\section{Parametrized Relativistic Classical Mechanics}
\setcounter{equation}{0}
Making the canonical formalism covariant means that the physical
laws, expressed by Poisson bracket relations, have to be invariant
under a transformation from one inertial system into another
inertial system. But transformations preserving the Poisson
brackets are canonical transformations. So a canonical system is
relativistic invariant if we have a canonical realization of the
relativity group. Manifest covariance means that in addition to
the requirement of relativistic invariance of the physical laws
the labeled trajectory of a particle in configuration space $\vec{q}(t)$
has to behave like a world line. This means that the relativity
postulate leads only to the requirement of a Poisson bracket
realization of the Poincare group, while manifest covariance
requires that the dynamical quantities $(t,\vec{q}(t))$ constitute an
space-time event \cite{Sudarshan}. There are now two approaches to
a manifest covariant extension of the canonical formalism in
classical mechanics. The first approach is that one describes the
particles by their canonical coordinates and the time coordinate
and then derives conditions that describe the fact that $(t,\vec{q}(t))$
transforms like an event in space-time. These additional
conditions lead then to the consequence that no interactions are
allowed \cite{Sudarshan}.

The alternative method that we will follow here is to use a
parameter formalism. In this approach the events that constitute
the world lines are labeled by an observer independent parameter
$s$ that increases monotonically as the world line is traversed.
The four space-time coordinates of an event on the world line are
then functions of this parameter: $x^{\mu}(s)$. Going now from one
inertial system to another one does not change the parameter:
$x^{\mu}(s)\rightarrow
{x'}^{\mu}(s)=\Lambda^{\mu}_{\;\nu}x^{\nu}(s)$. The four functions
$x^{\mu}(s)$ are regarded as the dynamical quantities, while the
parameter $s$ describes the evolution of the system. So the time
has no longer the two roles of a dynamical variable and an
evolution parameter.

It is now straight forward to develop a parametrized relativistic
mechanics \cite{Fanchi3}. One defines therefore a parameter-dependent action
\begin{equation}
S=\int_{s_1}^{s_2}ds\,L_s(q^{\mu},\mbox{\it\r{
q}}^{\,\mu},s),
\end{equation}
where $\mbox{\it\r{q}}^{\,\mu}$ is the derivation with respect to
the parameter $s$:
\begin{equation}
\mbox{\it\r{q}}^{\,\mu}=\frac{dq^{\mu}}{ds}.
\end{equation}
Requiring that the variation of the action vanishes: $\delta S=0$
leads to the parametrized version of the Euler-Lagrange equation:
\begin{equation}
\frac{d}{ds}\frac{\partial
L_s}{\partial\mbox{\it\r{q}}^{\,\mu}}-\frac{\partial L_s}{\partial
q^{\mu}}=0.
\end{equation}
With the Legendre transformation
\begin{equation}
K(q^{\mu},p_{\mu},s)=\mbox{\it\r{q}}^{\,\mu}p_{\mu}
-L_s(q^{\mu},\mbox{\it\r{q}}^{\,\mu},s)
\end{equation}
one then obtains the parametrized Hamilton equations:
\begin{equation}
\mbox{\it\r{q}}^{\,\mu}=\frac{\partial K}{\partial p_{\mu}}
\qquad\mathrm{and}\qquad \mbox{\it\r{p}}_{\mu}=-\frac{\partial
K}{\partial q^{\mu}}. \label{vierHamil}
\end{equation}
Using the Hamilton equations to calculate
\begin{equation}
\frac{d}{ds}f(q^{\mu},p_{\mu},s)=\{f,K\}_{PB}+\frac{\partial
f}{\partial s}
\end{equation}
one arrives at the four-space Poisson bracket
\begin{equation}
\{f,g\}_{PB}=\frac{\partial f}{\partial q^{\mu}}\frac{\partial
g}{\partial p_{\mu}}-\frac{\partial g}{\partial
q_{\mu}}\frac{\partial f}{\partial p^{\mu}},\label{4PB}
\end{equation}
for which follows
\begin{equation}
\{q^{\mu},p_{\nu}\}_{PB}=\delta_{\nu}^{\mu}\qquad\mathrm{and}\qquad
\{q^{\mu},q^{\nu}\}_{PB}=\{p_{\mu},p_{\nu}\}_{PB}=0.
\label{canova4PB}
\end{equation}

For example the covariant Hamiltonian of the free particle is
\begin{equation}
K=\frac{\eta^{\mu\nu}}{2m}p_{\mu}p_{\nu}\label{Kfrei}
\end{equation}
so that the Hamilton equations (\ref{vierHamil}) lead to
\begin{eqnarray}
&&\mbox{\it\r{p}}^{\,\mu}=0\;\;\,\quad\Rightarrow\quad
p_{\mu}=p_{0\mu}=\mathrm{const}\nonumber\\
\mathrm{and}&&\mbox{\it\r{q}}^{\,\mu}=\frac{p^{\mu}}{m}
\quad\Rightarrow\quad q^{\mu}=q_0^{\mu}+\frac{p_0^{\mu}}{m}s.
\end{eqnarray}
Variation of $q^{\mu}$ gives then $\delta q^{\mu}\delta q_{\mu}=
\frac{p_0^{\mu}p_{0\mu}}{m^2}(\delta s)^2=(\delta s)^2$ with the
initial condition $m^2=p_0^{\mu}p_{0\mu}$, which shows that the
parameter $s$ is just the proper time.

In the case of a charged particle in an electromagnetic field the
Hamiltonian (\ref{Kfrei}) generalizes to
\begin{equation}
K=\frac{\eta^{\mu\nu}}{2m}\left[p_{\mu}-eA_{\mu}\right]
\left[p_{\nu}-eA_{\nu}\right]=\frac{1}{2m}\pi^{\mu}\pi_{\mu},
\label{KFeld}
\end{equation}
with the kinetic momentum $\pi_{\mu}=p_{\mu}-eA_{\mu}$. The
Hamilton equations (\ref{vierHamil}) lead to
\begin{equation}
\mbox{\it\r{q}}_{\mu}=\frac{\pi_{\mu}}{m} \qquad\mathrm{and}\qquad
\mbox{\it\r{p}}_{\mu}=\frac{e}{m}\pi^{\nu}\partial_{\mu}A_{\nu}.
\label{Hamifrei}
\end{equation}
Combining these two equations gives $\mbox{\it\r{p}}_{\mu}
=e\mbox{\it\r{q}}^{\,\nu}\partial_{\mu}A_{\nu}$ and for the
derivation of the kinetic momentum with respect to $s$ one obtains
$\mbox{\it\r{$\!\!\!\!\pi$}}_{\mu}=\mbox{\it\r{p}}_{\mu}
-e\partial_{\nu}A_{\mu}\mbox{\it\r{q}}^{\,\nu}$. Equating then the
expressions for $\mbox{\it\r{p}}_{\mu}$ gives the Lorentz force
law
\begin{equation}
\mbox{\it\r{$\!\!\!\!\pi$}}_{\mu}=eF_{\mu\nu}\mbox{\it\r{q}}^{\,\nu}.
\label{4Lorentz}
\end{equation}
The classical mass is then a constant associated to the kinetic
momentum which can be obtained as follows. With (\ref{4Lorentz})
and (\ref{Hamifrei}) one can calculate
\begin{equation}
\mbox{\it\r{$\!\!\!\!\pi$}}_{\mu}\pi^{\mu}=\frac{1}{2}\frac{d}{ds}
\left(\pi_{\mu}\pi^{\mu}\right)=emF_{\mu\nu}\mbox{\it\r{q}}^{\,\nu}
\mbox{\it\r{q}}^{\,\mu}=0.
\end{equation}
From $\frac{d}{ds}\left(\pi_{\mu}\pi^{\mu}\right)=0$ follows then
that $\pi_{\mu}\pi^{\mu}=\pi_{0\mu}\pi_0^{\mu}$ is an integration
constant with respect to $s$. In order to be consistent with the
case $A_{\mu}=0$, where $p_{\mu}p^{\mu}=p_{0\mu}p_0^{\mu}=m^2$ one
chooses the integration constant as $\pi_{0\mu}\pi_0^{\mu}=m^2$.
This shows that the classical mass is a secondary concept in the
proper time formalism, while energy and momentum are primary
concepts.

\section{Deformation Quantization of Parametrized Classical Mechanics}
\setcounter{equation}{0}
Just as in the nonrelativistic case the connection of the four
dimensional Poisson bracket (\ref{4PB}) and the four dimensional
star product (\ref{comopro}) is given by
\begin{equation}
\lim_{\hbar\rightarrow
0}\frac{1}{\iu\hbar}\left[f,g\right]_{\starM}=\{f,g\}_{PB},
\end{equation}
so that the star commutators of the canonical coordinates are
\begin{equation}
\left[q^{\mu},p_{\nu}\right]_{\starM}=\iu\hbar\delta_{\nu}^{\mu}
\qquad\mathrm{and}\qquad \left[q^{\mu},q^{\nu}\right]_{\starM}=
\left[p_{\mu},p_{\nu}\right]_{\starM}=0.
\end{equation}
The structures of deformation quantization in the nonrelativistic
case can then be generalized to the four dimensional case in a 
straight forward manner. The development of the system in $s$ 
is generated by the four dimensional Hamiltonian. In the star 
product formalism this is described by the star exponential, 
which is in the four dimensional case given by
\begin{equation}
\mathrm{Exp}_{M}(Ks)=e_{\starM}^{-\iu sK/\hbar}=\sum_{n=0}^{\infty}
\frac{1}{n!}\left(\frac{-\iu s}{\hbar}\right)^n K^{n\starM},
\end{equation}
where $K^{n\starM}$ is the $n$-fold star product. The star
exponential fulfills the proper time generalization of the time
dependent Schr\"odinger equation:
\begin{equation}
\iu\hbar\frac{d}{ds}\mathrm{Exp}_{M}(Ks)=K\starM
\mathrm{Exp}_{M}(Ks).
\end{equation}
The calculations to determine the spectrum and the Wigner 
eigenfunctions then parallels the calculations in the non-relativistic case. 

But there is also an additional effect, because combining the Moyal 
product (\ref{comopro}) and the Clifford product (\ref{Cliffpro}) into 
one supersymmetric formalism one obtains a noncommutative version 
of space-time algebra. In the commutative or classical case the 
generalized Hamiltonian (\ref{KFeld}) can be written as
\begin{equation}
K=\frac{1}{2m}\fett{\pi}\starP\fett{\pi}=\frac{1}{2m}\fett{\pi}\cdot
\fett{\pi}
\end{equation}
with $\fett{\pi}=\pi^{\mu}\fett{\gamma}_{\mu}$. But if one introduces
noncommutativity via the Moyal product, the Moyal product of $\pi_{\mu}$
and $\pi_{\nu}$ is in general not symmetric in the indices, one rather
has
\begin{equation}
\left[\pi_{\mu},\pi_{\nu}\right]_{\starM}=\iu\hbar eF_{\mu\nu}.
\end{equation}
This leads then to the appearance of an additional term that describes
the spin:
\begin{equation}
K=\frac{1}{2m}\fett{\pi}\starMP\fett{\pi}=\frac{1}{2m}\left(\pi^{\mu}
\starM\pi^{\nu}\right)\left(\fett{\gamma}_{\mu}\starP\fett{\gamma}_{\nu}
\right)=\frac{1}{2m}\pi^{\mu}\pi_{\mu}
+\frac{1}{2m}\left(\pi^{\mu}\starM\pi^{\nu}\right) 
\fett{\gamma}_{\mu}\fett{\gamma}_{\nu}.\label{KFeld2}
\end{equation}
Such multivector valued additional terms due to noncommutativity 
also appear in the non-relativistic case \cite{Deform6} and on the 
phase space \cite{Susy}. In the case of a stationary particle 
in a homogenous magnetic field (\ref{KFeld2}) reduces to
\begin{equation}
K=-\frac{m}{2}+\iu\frac{e\hbar}{2m}B_3\fett{\gamma}_1\fett{\gamma}_2,
\end{equation}
so that one has the spin eigenfunctions $\frac{1}{2}\pm\frac{\iu}{2}
\fett{\gamma}_1\fett{\gamma}_2$, that fulfil
\begin{equation}
\iu\fett{\gamma}_1\fett{\gamma}_2\starP\left(
\frac{1}{2}\pm\frac{\iu}{2}\fett{\gamma}_1\fett{\gamma}_2\right)
=\pm\left(\frac{1}{2}\pm\frac{\iu}{2}\fett{\gamma}_1\fett{\gamma}_2\right).
\end{equation}

\section{Conclusions}
\setcounter{equation}{0}

\qquad The algebra of the gamma matrices and the operator algebra can
both be described in the star product formalism. So combining the two
star products allows to formally describe Dirac theory. There are 
nevertheless two problems that directly arise in this context. On the one
hand the resulting formalism is not supersymmetric, because one uses
a three dimensional bosonic star product and a four dimensional fermionic
star product. On the other hand Dirac theory does not appear as the
deformation of a classical relativistic theory. Both problems can be 
solved if one applies deformation quantization to parametrized relativistic
theory. So besides the physical arguments in favour of parametrized
relativistic theory deformation quantization also gives a formal 
preference. Moreover the combination of the bosonic and the fermionic
star products describe a noncommutative version of geometric algebra   
that produces spin terms automatically.



\begin{thebibliography}{9999}

\bibitem{Gerstenhaber} M. Gerstenhaber,
Ann. Math. {\bf 79} (1964) 59.

\bibitem{Bayen} F. Bayen, M. Flato, C. Fronsdal, A. Lichnerowicz and D.
Sternheimer,
Ann. Phys. {\bf 111} (1978) 61, 111.

\bibitem{Deform3} A. C. Hirshfeld and P. Henselder,
Ann. Phys. {\bf 302} (2002) 59.

\bibitem{Deform5} A.\,C.\,Hirshfeld, P.\,Henselder and T.\,Spernat,
Ann. Phys. {\bf 314} (2004) 75. 

\bibitem{Berezin1} F.\,A.\,Berezin and M.\,S.\,Marinov,
Ann. Phys. {\bf 104} (1977) 336.

\bibitem{DoranGull} A.\ Lasenby, C.\ Doran and  S.\ Gull, 
J. Math. Phys. {\bf 34} (1993) 3683. 

\bibitem{Doran1} C.\ Doran, A.\ Lasenby,
Geometric Algebra for Physicists,
Cambridge University Press, 2003.

\bibitem{Deform6} P.\ Henselder, A.\ C.\ Hirshfeld, T.\ Spernat,
Ann. Phys. {\bf 317} (2005) 107.

\bibitem{Hestenes1} D.\,Hestenes,
Space-Time Algebra, Gordon and Breach, 1966.

\bibitem{Doran}  C.\ Doran, A.\ Lasenby, S.\ Gull, S.\ Somaroo 
and A.\ Challinor,
{\it Spacetime Algebra and Electron Physics}
in P. W. Hawkes (Ed.), Advances in Imaging and Electron Physics, 
Vol. 95, p. 271-386 (Academic Press, 1996). 

\bibitem{Fanchi1} J.\ R.\ Fanchi, 
Am. J. Phys. {\bf 49} (1981) 850.

\bibitem{Bolivar} A.\ O.\ Bolivar,
J. Math. Phys. {\bf 42} (2001) 4020.

\bibitem{Liang} M.\ Liang and Y.\ Sun,
Ann. Phys. {\bf 314} (2004) 1.

\bibitem{Fanchi2} J.\ R.\ Fanchi,
Found. Phys. {\bf 23} (1993) 487.

\bibitem{Fanchi3} J.\ R.\ Fanchi,
Parametrized Relativistic Quantum Theory, 
Kluwer Academic, Dordrecht, 1993.     

\bibitem{Johns} O.\ D.\ Johns, 
Analytical Mechanics for Relativity and Quantum Mechanics, 
Oxford University Press, 2005.

\bibitem{Pavsic} M.\ Pavsic, 
Found. Phys. {\bf 31} (2001) 1185.

\bibitem{Susy} P.\ Henselder, Phys. Lett. A {\bf 363} (2007) 378.

\bibitem{Galapon} E.\ A.\ Galapon, 
What could have we been missing while Pauli's Theorem was in force?,
quant-ph/0303106. 

\bibitem{Sudarshan} E.\,C.\,G.\,Sudarshan and N.\,Mukunda,
"Classical Dynamics: A Modern Perspective", John Wiley \& Sons
(1974).




\end{thebibliography}
\end{document}